%% file: main.tex
\documentclass[10pt,twocolumn,letterpaper]{article}

\usepackage{cvpr}
\usepackage{times}
\usepackage{epsfig}
\usepackage{graphicx}
\usepackage{amsmath}
\usepackage{amssymb}

\usepackage[breaklinks=true,bookmarks=false]{hyperref}

\cvprfinalcopy 

\begin{document}
\setlength{\parskip}{0pt}

\title{What is the right Addressing scheme for India?}

\author{Dr. Kabir Rustogi\\
\and
Dr. Santanu Bhattacharya\\
\and
Margaret Church\\
\and
Dr. Ramesh Raskar
}

\maketitle

\begin{abstract}

    Computer generated addresses are coming to your neighborhood because most places in the world do not have an assigned meaningful street address. In India, 80\% of the addresses are written with respect to a landmark which typically lies between 50-1500 meters of the actual address; such addresses make geolocating very challenging. Accuracy in geolocation is critical for emergency services to navigate quickly to reach you and for logistics industries to improve on-time performance and efficient routing of the package coming to your house. In this paper, we explore suggested addressing schemes for India, to determine what use cases and potential technologies will have the best adoption and therefore, greatest impact. 
\end{abstract}

\section{Introduction}
\input{intro}

\section{Cost of bad addresses}
\input{cost}

\section{Overview of current Geolocating Services}
\input{overview}

\section{Attributes of an Ideal Machine Code}
\input{attributes}

\section{Conclusion: Lessons from Aadhaar}
\input{conclusion}

\section{About the Authors}
\input{about}

{\small
\bibliographystyle{ieee}
\bibliography{egbib}
}

\end{document}

%% file: intro.tex
Zippr, eLoc and beyond .. which system will work for India? \par
Currently there is a rush to use machine generated codes such as 4ZXR3B (eLoc) or CAFE0098 (Zippr). These methods have proven to work in a few ways, but such systems can be confusing for the adoptee and there are technical drawbacks as well. It is critical that India adopts the most effective scheme, and not the scheme that is most readily available or has the largest company behind it. We ask: What are the requirements for machine codes so that they are easy for a layman, easy for a service company (eCommerce, taxi etc) and suitable for computer systems? \par
Here we review the desired features, compare various solutions, and suggest a path for widespread adoption of machine codes in India.
\begin{figure}[t]
\begin{center}
   \includegraphics[width=0.9\linewidth]{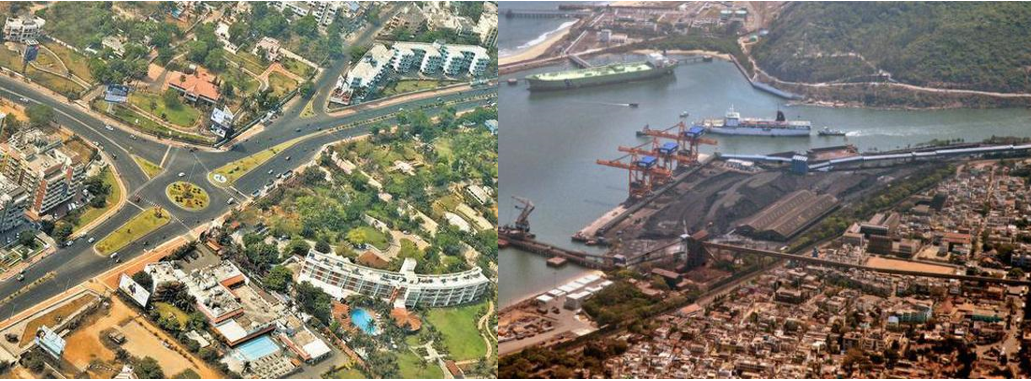}
\end{center}
   \caption{Indian cities are full of contrast; Visakhapatnam (erstwhile “Vizag”) has a well laid-out city center with proper addresses while the port area is overgrown and chaotic.}
\label{fig:intro}
\end{figure}

%% file: cost.tex
The economic impact of bad addresses in India is significant: our estimate from the top industries indicate that \textbf{poor addresses cost India \$10-14B annually, ~0.5\% of the GDP}; see Appendix 1 for details. Addresses we have encountered contain local abbreviations, colloquial neighborhood names, points-of-interest, embedded unclear directions and variations due to local languages being transliterated in English for writing. Localities and pin-codes have poor localization. In India, the average area covered by localities and pin-codes is around 1.5 and 179 square kilometers, respectively, where the latter may contain up to a million households. To make things worse, 20-30\% of written pin-codes are incorrect. The average distance between the location of a point-of-interest written in an address and the actual location of the doorstep is around 400m, implying that including landmarks in the address does not significantly improve resolution. Moreover, landmarks are difficult to use for geolocating because there are about 10 million points-of-interest (e.g., “State Bank ATM”) in just the top 200 Indian cities, making it a complex cataloguing and updating task, especially in rapidly changing smaller towns.

%% file: overview.tex
Many systems have been proposed in the past few years to solve the problem of extracting precise geocodes from addresses. They primarily fall under two categories.

\subsection{Hardcoded Addresses: Disambiguation and Geocoding}

Structured hardcoded address, e.g., “26 Gandhi Road, Dhule, Maharashtra, India” can be converted to a fairly precise geocode. Unfortunately, only about 30\% of address are written in such format in India. A more common address in the same town would be “B56 Niman Nagar, Near Green Park, Dhule, Maharashtra, India”.


\begin{table*}[t]
\begin{center}
\includegraphics[width=0.9\textwidth]{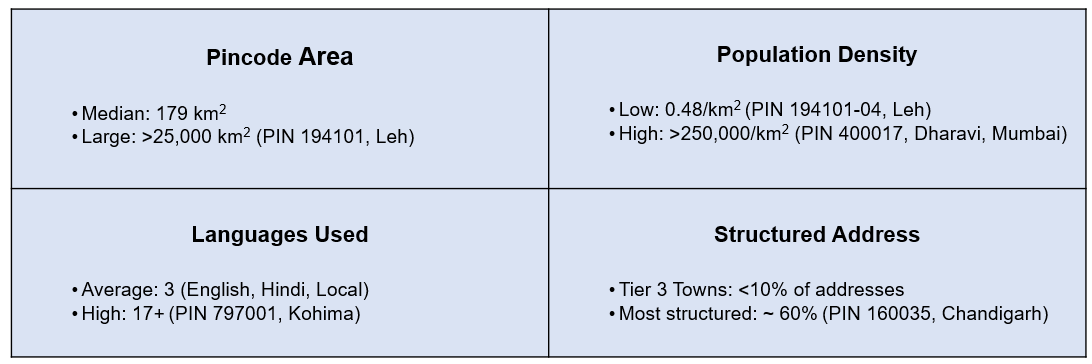}
\end{center}
   \caption{Large pin-code areas, dense population, lack of structure and multilingual support for addresses written and transliterated to English makes address disambiguation a complex task.}
\label{fig:overview1}
\end{table*}


For addresses that have structures, companies have developed algorithms to disambiguate a raw address string into a proper format, typically consisting of features such as state, city, locality, sublocality, street and house number. These features are then associated with geocodes or polygon boundaries, which are either obtained by crowdsourcing or by some form of surveying. Examples of such services include Google Maps, MapMyIndia, Delhivery’s AddFix (not publicly available, see \cite{1}). This approach proves effective for resolving an address to an order of a few hundred meters, but rarely down to the house number. Its effectiveness is also limited by how structured the input address is and the depth of geospatial data available for each locality feature. Hence, for India this is not a scalable approach and for 70\% of sites with no street names, there is no easy solution.

\subsection{Machine Coded Addresses}

A more disruptive way to solve the above problem is to replace traditional addresses using a machine generated code for each location. You can imagine an ‘Aadhaar\cite{7}-like code for each address’. Aadhaar is a 12-digit random identification number issued by the Unique Identification Authority of India (UIDAI) on behalf of the Government to the residents of India for the purpose of establishing the unique identity of every single person in the country. Machine codes promise to be easily readable, are fast and are available through APIs for programmatic integration. Recently, there has been a wave of services that aim to achieve this goal. Machine codes can be classified into three categories:

\begin{enumerate}
    \item Short-codes: Machine generated codes are assigned to each unique address record. This is often achieved by manual surveying or crowdsourcing the location of each address. Examples include:
    \begin{itemize}
        \item \textbf{eLoc} - randomized alphanumeric code for each known address record, e.g., DIO5L6; see \cite{2}
        \item \textbf{Zippr} -  customizable alphanumeric code available for any point on a map, e.g., CAFE0098; see \cite{3}
    \end{itemize}
    \item Auto-codes: This system follows a grid-based approach. The Earth is divided into 3m x 3m imaginary grids and a code is automatically generated for each grid. Given any latitude and longitude (lat-long), an algorithm identifies the grid it belongs to and returns the corresponding auto-code. Examples include:
    \begin{itemize}
        \item \textbf{Plus Codes} - alphanumeric codes available for each 3m x 3m area on a map, e.g., 7JWVF23W+GQQ; see \cite{4};
        \item \textbf{What3Words (W3W)} - collection of 3 random English words available for each 3m x 3m area, e.g., parrot.casino.failed; see \cite{5};
    \end{itemize}
    
    \item Street-codes: This system follows a street-based approach. Each point on a map is assigned a street number based on its distance in meters from the southwest corner of the nearest street. The system uses a street name if it already exists. If the street has no name, the system creates a short street name using north-south-east-west orientation with respect to the city or town center. Examples include:
    \begin{itemize}
        \item \textbf{Robocodes} - four fields with hierarchical and linear descriptors, namely position with respect to a street, locality/street name, city, state/country, e.g., 90C.NE88.Dhule.MhIn; in this case, the site is on the 88th unnamed street north of Dhule city, and the location is 90 meters from the southwest corner of this street; see \cite{6}.
    \end{itemize}
\end{enumerate}
The combinatorics behind the above coding schemes can be found in Appendix 2. Each of the above services has its own advantages and disadvantages, which have an impact on its adoption among people.

\begin{figure}[t]
\begin{center}
   \includegraphics[width=0.9\linewidth]{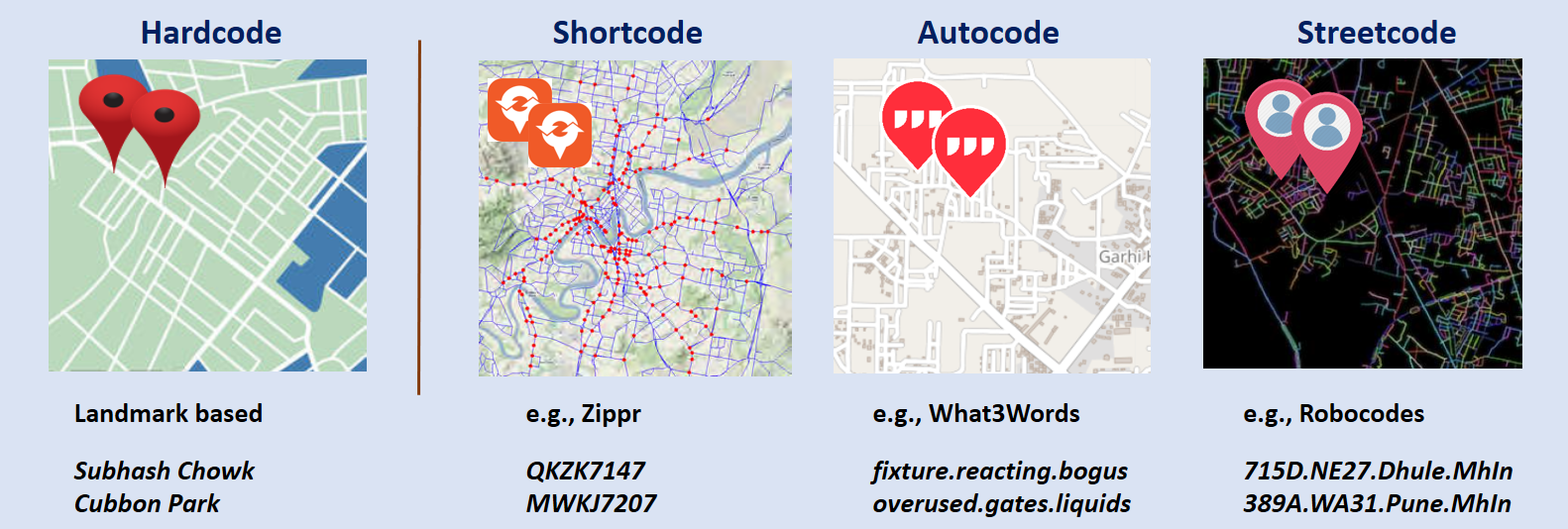}
\end{center}
   \caption{Evolution of geocoding}
\label{fig:overview2}
\end{figure}


%% file: attributes.tex
Most of the machine codes described above are built as engineering services and often ignore the human aspects that will ensure their early adoption and ease of use for the public.

\subsection{Are they Memorable?}

Machine codes should have an easy recall among its user base. In this respect, Plus Codes and eLoc do not fare very well because they use seemingly random alphanumeric codes. Imagine giving your pizza delivery person your Plus Code 7JWVF36Q+P4 over the phone. Zippr and W3W are somewhat memorable, but still not scalable enough in the long run, since they do not contain any spatial clue or relation to the actual address. As a result, two neighbors may have completely different codes.

\begin{figure}[t]
\begin{center}
   \includegraphics[width=0.9\linewidth]{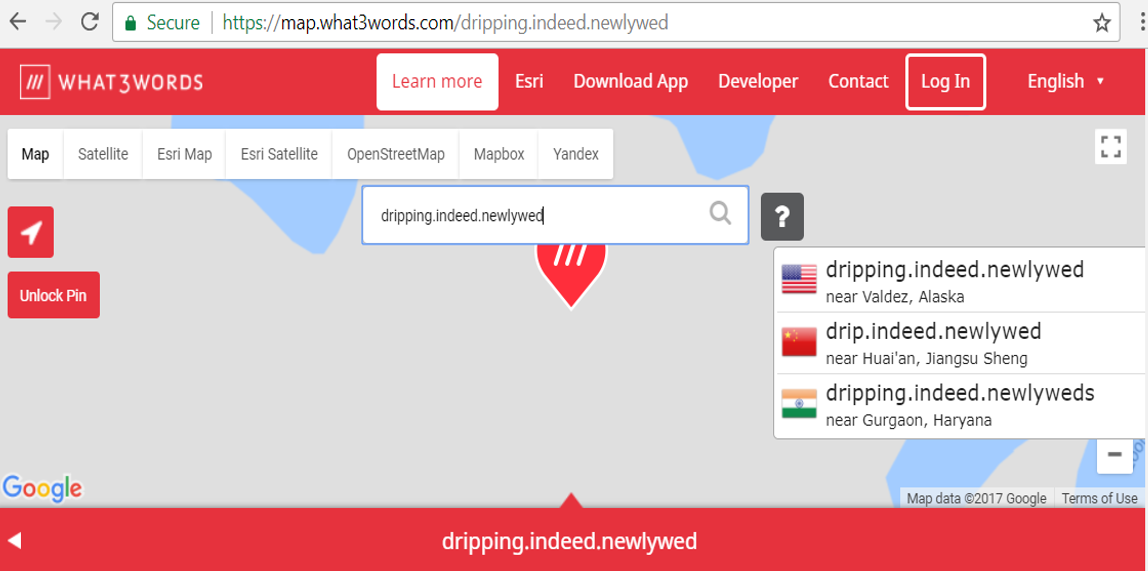}
\end{center}
   \caption{A slight variation of the What3Word code dripping.indeed.newlywed can take you over 13,000 km away from Valdez, Alaska, USA to Gurgaon, Haryana, India.}
\label{fig:attributes1}
\end{figure}


Moreover, What3Words can be quite confusing to use in countries where English is not the first language. Some of the words may be quite uncommon for the common man and minor differences in the word may lead to completely different locations on the globe; see Figure 3 for an example. Robocodes are promising in this respect, since they include features that people are already familiar with and are used to writing in their addresses, e.g. street, locality and city/state names.

\subsection{Are they Intuitive?}

Addresses have evolved over the millennia and have the following inherent properties that help us understand and quickly infer their relationship with other addresses:

\begin{itemize}
    \item \textbf{Hierarchy}: Typically, addresses contain features in decreasing level of granularity, from rooftop name to state/country name. This helps us identify if two addresses are in the same locality/city/state/country.
    \item \textbf{Linearity/Continuity}: Typically, addresses which are close to each other have related names that one can understand intuitively. This helps us identify how distant two addresses are on the same street.
\end{itemize}

\begin{figure}[t]
\begin{center}
   \includegraphics[width=0.9\linewidth]{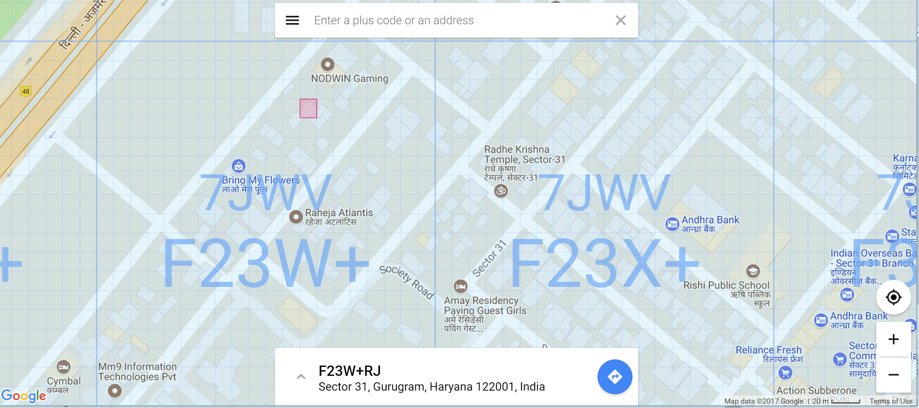}
\end{center}
   \caption{An illustration of the grid-based approach that powers Plus Codes.}
\label{fig:attributes2}
\end{figure}

    
Most geocoding services such as W3W, Zippr, eLoc remove the geometric relations between addresses, blocking human intuition to naturally understand them. They neither follow a hierarchical system, nor are continuous, hence making it impossible for people to derive them logically. Plus Codes are somewhat more promising in this respect. They are generated by dividing the earth into grids of 20 x 20 degrees, and then further dividing each grid into smaller grids, all the way down to a 3m resolution. Plus Codes assign an alphanumeric value to grids of different sizes, thereby incorporating the concept of hierarchy and linearity in the schema.

While the combinatorics presented in Appendix 2 indicate that alphanumeric short-codes must comprise of a minimum of 6 characters to uniquely identify every associated address, it is advisable to have more characters in the code. This allows the code to incorporate desirable attributes such as being memorable, e.g. CAFE0048, or being hierarchical, and as a result becomes more intuitive. For instance, alphanumeric auto-codes must comprise of a minimum of 9 characters to uniquely identify every point on the Earth’s surface. However,  Plus Codes are made up of 11 characters, so that they can incorporate a sense of hierarchy in the code, e.g., each point in Delhi begins with the characters 7JW; see Figure 4 for illustration.

Robocodes further refine this approach to make it even more intuitive for the user. Instead of using an imaginary grid based system, they use actual street names and locality/city/state names in its nomenclature, to emphasize on hierarchical and linear relationships.

\begin{figure}[t]
\begin{center}
   \includegraphics[width=0.9\linewidth]{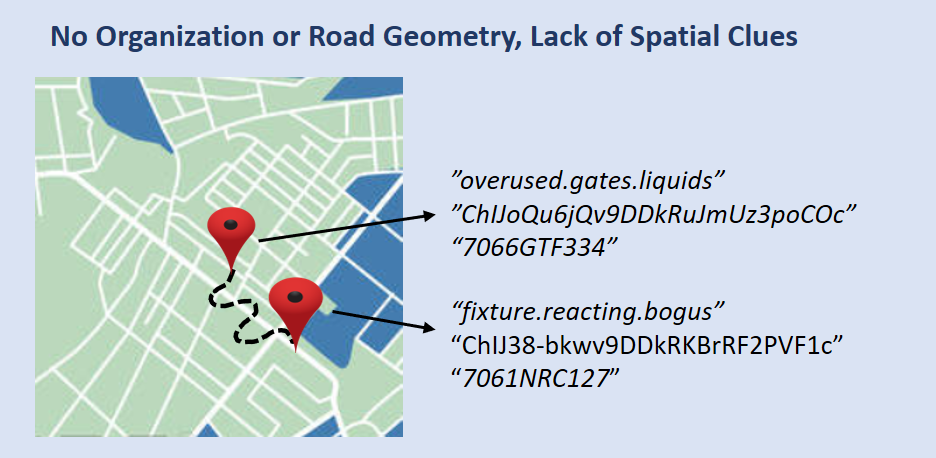}
\end{center}
   \caption{Many auto-codes do not provide spatial clues, structural organization or usage of road geometry to make the codes intuitive for its users.}
\label{fig:attributes3}
\end{figure}


\subsection{Ease of Adoption? Backward Compatible?}

A key factor for the success of any geocoding service is the ease with which a user can convert their address to a machine code. Most of available services require the user to input a lat-long to generate a code. However, to capture precise lat-longs from a user is difficult.

The easiest way to get lat-longs is by capturing the GPS coordinates of the current location from the user’s mobile device, but this approach poses challenges. Sometimes the user may not be present at the location for which the lat-long is required, e.g., a parent ordering food for children at home while they are in office or informing a taxi app where you want to go. Moreover, the lat-longs captured by the devices are not always accurate. From a sample of 500,000 locations captured by a Delhivery, an Indian eCommerce logistics company, only 50\% had a reported accuracy of within 50 meters and only 10\% were within 5 meters. Poor GPS signal in densely built areas, low quality of GPS receptors in budget phones, users keeping their GPS off to conserve battery, etc. contribute to inaccuracy.

Another way to obtain lat-longs is by providing an interface to the user where they can mark their location on a map. This approach ensures that device errors do not play a role; however, human errors may be far more damaging. Delhivery piloted this approach with a sample of eCommerce customers. It was reported that only 25\% of the customers were able to mark their location within 100m of the location captured by the ground staff subsequently. Unfamiliarity with digital maps, the inconvenience of performing an extra step for the user, etc. contributed to the errors.

A third approach could be to manually geotag address records by employing extensive surveying teams. Recently, the state government of Andhra Pradesh commissioned Zippr to manually tag each and every household in the state, so that they would be onboarded to the short-code platform; see \cite{3}.

To a large extent, Robocodes can solve this problem, due to their linear and intuitive design. People can interpolate and infer from their neighbor’s robocode, since street name is shared and street number is sequential. Although this approach is somewhat limited due to its dependency on at least someone in the street to have the correct robocode of their address, it has the potential of spreading organically. For example, if the locations ‘200 Road N12’ and ‘220 Road N12’ start using the code to realize benefits, e.g., faster taxi arrivals, then the people between \#200 and \#220 may start using the robocodes with intermediate numbers.

The key for wide adoption also lies in whether a geocoding scheme can be used without complex technologies. Can people locate an address without using a smartphone, e.g., by simply using a printed map or a billboard? When addresses can be interpolated, it becomes easier to map a whole town with only a few street names and region names.

\subsection{Are they Future Proof?}

What happens when new streets emerge? Or if one wants to support locations inside a national park or in middle of the ocean? Grid-aligned auto-codes easily support this, but not short-codes, such as Zippr. Robocodes behave like new domain names or IP addresses emerging on the internet. They require someone to include new roads and will require updates from trusted agencies to spread.

As cities expand vertically and transportation evolves, it will be necessary to capture altitude in the machine codes. None of the mainstream solutions today provide this feature easily.

\subsection{Machine Friendly? Free? Open Source?}

Most of the services are built so that a machine can easily read the codes and convert them to a standard lat-long for further geospatial computations. The consumers of this service, e.g., enterprises who want to perform route optimization, are able to perform transformations on the machine codes by calling the APIs built by their creators. There are several factors to consider: Does the scheme require live internet connectivity for the API conversion call? Does it need a smartphone or can it be shared via SMS? Does it cost an intermediate business to make these API calls? Can the underlying lookup table be downloaded for local use? How easy is it to create and populate these codes? Who maintains them?

\begin{table*}
\begin{center}
\includegraphics[width=0.9\textwidth]{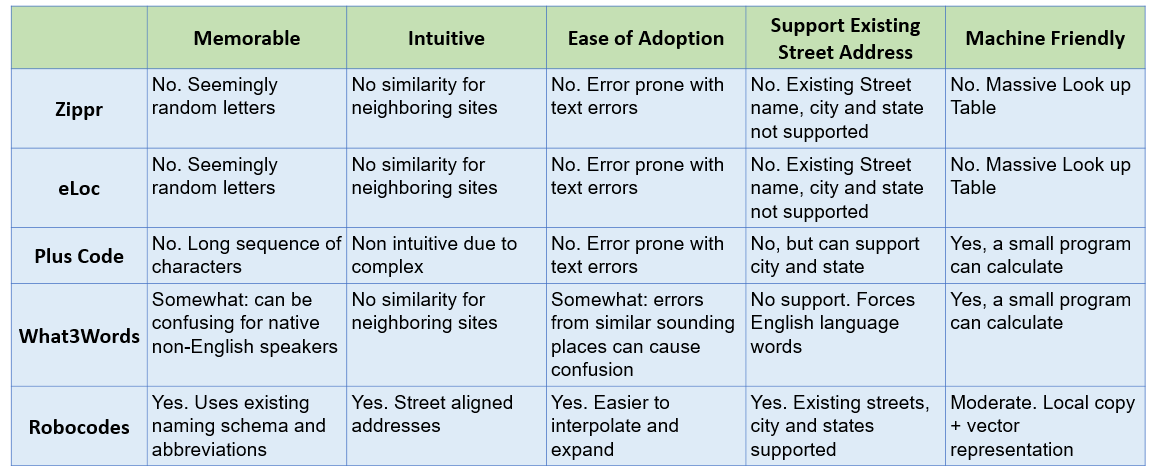}
\end{center}
   \caption{Comparison of popular geocoding systems}
\label{fig:attributes4}
\end{table*}


Auto-codes, such as W3W and Plus Codes provide a clever solution. The auto-codes can be calculated from a lat-long on a local device, without internet access, using a very small program that has about 40,000 dictionary words or purely mathematical formulas. Short-codes such as eLoc and Zippr, need a massive lookup table with no ability to interpolate or validate locally as neighboring locations have different codes. Most of the APIs, e.g., W3W, eLoc are opaque and are relatively expensive for a country like India. For example, Google APIs cost \$1000 for 2 million geocoding queries; a relatively large sum for a small business in India. A small digital player with 10 million users that make only 2 map queries a day will incur \$3M in API charges. A nationwide business will have to pay 100’s of millions of dollars per year to these companies if they dominate, become monopolies or locate outside India to avoid possible price regulations. This can be a real threat to digital businesses.

Robocodes are free, open source and could be communicated using SMS because they are human-friendly and are backward compatible, i.e., existing street addresses can be kept as they are. However, Robocodes require significant processing to generate in the first place. Hence, Robocodes require some entity to maintain tamper-proof versions and update as and when new street geometries are identified.

%% file: conclusion.tex
The need to create an addressing scheme for India is evident based on the rush of startups and companies in this space. This is the new “Aadhaar” mindset, a code for each location. However, this approach is faulty, as an ID for a person is a very different from an ID for a location.

A place is understood for navigation or for mental anchoring via hierarchy, linearity and memorability. A user-ID like Aadhaar is intentionally obscure and should not be based on any identifiable parameters. Addresses, on the other hand, should be identifiable and similar to other addresses in the neighborhood. Addresses are shared verbally and need to become part of the culture for rapid adoption and must not be obscure. We believe that using seemingly random codes as address for locations is misdirected. Just as Aadhaar has made banking, mobile phone and government services accessible to the population at the bottom of the pyramid, a proper addressing scheme will be required for geolocated government services but \textit{only if the scheme is simple enough to be used by everyone}.

It is time for eCommerce, on-demand transport, package delivery, government agencies, hospitality and tourism sector to come together to work on an open source, human-friendly and business-friendly scheme. We hope our analysis of existing solutions and desired features will spawn a healthy debate and methodical approach towards the best addressing scheme for India.

We are concerned that a well-intentioned but misdirected effort to assign addresses will cause irreversible damage to the growth of digital economy in India. We already see experimentation with Zippr codes in Hyderabad and trials with eLoc in Delhi. A piecemeal approach by multiple states, without deeper analysis or comparison of products available, is going to create multiple solutions without interoperability; causing a huge burden downstream. A digital address is as important as a digital identify; now imagine if every state created their own Aadhaar?

%% file: about.tex
\textbf{Dr. Kabir Rustogi} leads the Data Science team at Delhivery, India’s largest eCommerce logistics company. A published author, he was previously a Senior Lecturer of Operations Research at The University of Greenwich, UK.

\textbf{Dr. Santanu Bhattacharya} is scientist collaborating with Camera Culture Group at MIT Media Lab. A serial entrepreneur who has led Emerging Market Phones at Facebook, he is a former physicist from NASA Goddard Space Flight Center.

\textbf{Margaret Church} is coordinator of the Emerging Worlds Initiative at MIT Media Lab.

\textbf{Dr. Ramesh Raskar} is Associate Professor at MIT Media Lab and leads the Emerging Worlds Initiative at MIT which aims to use global digital platforms to solve major social problems.

%% file: main.bbl
\begin{thebibliography}{1}

\bibitem{1}
K.~Rustogi.
\newblock {\em
  \href{https://medium.com/@kabirrustogi/learning-to-decode-unstructured-indian-addresses-c80ffcda2e84}{Learning
  to Decode Unstructured Indian Addresses}}.
\newblock Medium, 2017.

\bibitem{7}
{\em \href{https://uidai.gov.in/your-aadhaar/about-aadhaar.html}{About
  Aadhaar}}.
\newblock Government of India.

\bibitem{2}
{\em
  \href{https://economictimes.indiatimes.com/news/economy/policy/now-government-to-start-mapping-your-address-digitally/articleshow/61666640.cms}{Now,
  government to start mapping your address digitally}}.
\newblock The Economic Times, 2017.

\bibitem{3}
{\em
  \href{http://www.livemint.com/Politics/epYBSl0nVGaKa8wkx9zPWK/Andhra-Pradesh-kicks-off-Smart-Pulse-Survey-of-148-million.html}{Andhra
  Pradesh kicks off Smart Pulse Survey of 14.8 million households}}.
\newblock Live Mint, 2016.

\bibitem{4}
D.~Rinckes.
\newblock {\em
  \href{https://opensource.googleblog.com/2015/04/open-location-code-addresses-for.html}{Open
  Location Code: Addresses for everything, everywhere}}.
\newblock Google Open Source Blog, 2015.

\bibitem{5}
{\em
  \href{https://qz.com/705273/mongolia-is-changing-all-its-addresses-to-three-word-phrases/}{Mongolia
  is changing all its addresses to three-word phrases}}.
\newblock Quartz, 2016.

\bibitem{6}
Hughes F. Raj A. Tsourides K. Ravichandran D. Murthy S. Dhruv K. Garg S.
  Malhotra J. Doo B. Kermani~G. Demir, I. and Raskar R.
\newblock {\em \href{http://ieeexplore.ieee.org/document/8014926/}{Robocodes:
  Towards Generative Street Addresses from Satellite Imagery}}.
\newblock In Computer Vision and Pattern Recognition Workshops (CVPRW), 2017
  IEEE Conference on (pp. 1486-1495). IEEE., 2017.

\end{thebibliography}
